\documentclass[10pt]{amsart}

\usepackage{amsmath}
\usepackage{amssymb}
\usepackage{graphicx}
\usepackage{subfigure}
\usepackage{cite}

\title{Many-body dark solitons in one-dimensional hard-core Bose gases}
\author{Manuele Tettamanti $^{1,2}$ and Alberto Parola $^{3}$}
\address{%
$^{1}$ \quad Dipartimento di Fisica ``Giuseppe Occhialini", Universit\`a di Milano-Bicocca, Piazza della Scienza 3, 20126 Milano, Italy \\
$^{2}$ \quad INFN — Sezione di Milano-Bicocca, Piazza della Scienza 3, 20126 Milano, Italy \\
$^{3}$ \quad Dipartimento di Scienza e Alta Tecnologia, Universit\`a degli Studi dell’Insubria, Via Valleggio 11, 22100 Como, Italy}

\begin{document}

\begin{abstract}
The existence and stability of solitonic states in one-dimensional repulsive Bose-Einstein condensates is 
investigated within a fully many-body framework by considering the limit of infinite repulsion (Tonks-Girardeau gas). 
A class of stationary, shape-invariant states propagating at constant velocity are explicitly found and 
compared to the known solution of the Gross-Pitaevskii equation. The typical features attributed to nonlinearity are thus recovered in a purely linear theory, provided the full many-particle physics is correctly accounted for. However, the formation dynamics predicted by the Gross-Pitaevskii approximation considerably differs from the exact many-body evolution. 
\end{abstract}
\maketitle

\section{Introduction}
Solitons are non-uniform, stationary solutions of a translationally invariant, non-linear equation which can propagate with constant velocity (less than a critical value) without changing shape. Exact solitonic solutions are known in several frameworks, ranging from fluid-dynamics to optics, including the one-dimensional Gross-Pitaevskii equation (GPE) with cubic nonlinearity, describing the dynamics of a Bose-Einstein condensate (BEC) \cite{solitons}. Dark solitons are localized rarefaction in an asymptotically uniform fluid and they have been proven to exist and to be stable in the one-dimensional GPE with repulsive interactions \cite{pita}. By a Galileo transformation, a moving soliton in a fluid asymptotically at rest can be also seen as a soliton at rest in a flowing condensate. If the fluid is asymptotically at rest, the density profile of the dark soliton vanishes at the minimum, while if the fluid flows with a finite velocity the soliton becomes ``gray", i.e., its minimum density is non-zero. The maximum asymptotic fluid velocity supporting a solitonic solution coincides with the sound velocity in the asymptotic region.

In the laboratory, soliton-like structures have been created in BECs for more than 20 years \cite{burger,den} through techniques of phase imprinting, density engineering, and a combination of the two \cite{review}. Experimental solitons are clearly related to the exact solution of the GPE and their physical origin traces back to the balance between dispersion and self-interaction, which stabilizes a non-uniform profile. However, although the qualitative features of a soliton are clearly reproduced in experiments, it is difficult to ascertain the precise solitonic nature of the observed density dip: indeed, the underlying three dimensionality of the cigar-shaped traps, the finite lifetime of the condensate, the presence of longitudinal confinement breaking translational invariance and finite temperature effects are only a few unavoidable perturbations which must be taken into account when comparing analytical and experimental results \cite{review}. Solitonic states are often identified with a particular branch of excitation (dubbed ``Lieb type II'') of the exact Lieb-Liniger solution \cite{lieb,lieb2} of a $\delta$-interacting, one-dimensional Bose gas \cite{ll1,ll2,ll3,ll4,ll5}. However, growing evidence \cite{dynamics1,dynamics2,dynamics3,dynamics4} suggests that, starting from a superposition of these excited states, the exact dynamics drives particles into the density dip which eventually fades away. Solitons, thus, would emerge only as mean field stationary solutions which do not carry over to the full many-body Schr\"odinger equation. On the other hand, it has been also argued that a single quantum measurement of a quasiuniform many-body state may still display a well-defined density dip. After an average over many different measures, however, the dip is bound to vanish due to fluctuations of the soliton position \cite{misura1,misura2,misura3}. In any case, a well-defined, broken symmetry, stationary solution of the full many-body problem has not been clearly identified yet. In this article, we ask whether it is possible to determine and dynamically generate ``many-body solitons", i.e., exact stationary solutions of an interacting BEC in one dimension. 

Here we investigate the strong coupling limit of the Lieb-Liniger solution, the so-called Tonks-Girardeau (TG)
gas, whose eigenstates and dynamics exactly maps onto those of a free Fermi gas \cite{tg,bloch,weiss}. 
Confining our attention to the TG model is not a serious limitation: it is known that the strong-coupling regime 
of a one-dimensional, $\delta$-interacting Bose gas is not pathological, as opposed to the ideal gas limit. Its equilibrium properties, together with the spectrum of elementary excitations, are in fact smooth functions of the interaction strength all the way to the hard-core limit \cite{lieb,lieb2}. Most importantly, the long time dynamics of a one-dimensional, interacting Bose gas has been shown to be asymptotically described precisely by the TG model for any value of the (repulsive) coupling constant \cite{tgdyn2,tgdyn3}. An obvious consequence of the strict one dimensionality of our model is the absence of true off-diagonal long range order, i.e. the absence of a condensate and of macroscopic phase coherence \cite{pita}. Belonging to the class of Luttinger liquids, the TG gas still displays quasi-long range order, characterized by a density matrix decaying as $r^{-1/2}$ and by a divergent momentum distribution \cite{vaidya}. Therefore we believe that such a toy model may indeed serve as a useful guidance for the theoretical investigations of quasi one-dimensional Bose-Einstein Condensates, allowing, at the same time, for a full treatment of the many-body effects.

In this article, we show that a careful analysis of the solitonic GPE solution allows identification of the exact stationary states of the TG model representing a dark soliton propagating with constant velocity. A jump in the phase of the wavefunction goes along with the density modulation, in close analogy with the known GPE soliton. We also numerically investigate the possibility to dynamically generate TG solitons.

\vskip 1cm
\section{Many-body solution} 

The GPE with cubic nonlinearity is known to describe a Bose gas in the weak coupling regime and thus is not expected to faithfully reproduce the exact many-body dynamics of the strongly interacting limit. However, by replacing the cubic term with a {\it quintic} nonlinearity and fixing the coupling $g$ to the specific value $g=\frac{\hbar^2\pi^2}{2m}$, the exact low-energy excitation spectrum of the TG gas is recovered. Hence, this modified version of the GPE is usually adopted to represent a strongly interacting Bose gas in one dimension \cite{gp5,menotti} and its validity has thoroughly been tested \cite{mah}. In the following, we will make use of this equation, although the results obtained by means of the ordinary cubic GPE show no qualitative differences.

We start from the analytic, stationary solution of the quintic GPE representing a dark soliton in the co-moving frame  
(i.e., the reference frame at rest with the soliton, while the fluid velocity is fixed at infinity) which reads \cite{gp5,universe}: 

\begin{eqnarray}
\label{psi}
\Phi(x) &=& \sqrt{n(x)}\,e^{i\,\varphi(x)} \\
\frac{n(x)}{n_\infty} &=&  1-\frac{3\,\delta^2}{2+\sqrt{4-3\,\delta^2}\,\cosh\left ( 2\pi\,\delta\,n_\infty x\right)}
\label{solidens}
\\
\varphi(x) &=& -\pi\,\sqrt{1-\delta^2}\,n_\infty\,x - \nonumber \\
&& \frac{{\rm sign}\, x}{2}\,
\arccos \left [ \frac{1+\alpha \cosh\left ( 2\pi\,\delta\,n_\infty x\right)}{
\alpha + \cosh\left ( 2\pi\,\delta\,n_\infty x\right)} \right ] 
\label{soliphase}
\end{eqnarray}
where $\alpha = \frac{2-3\,\delta^2}{\sqrt{4-3\,\delta^2}}$ and the sign\,$x$ function is required by the symmetry $\varphi(-x)=-\varphi(x)$. The parameter $\delta$ is related to the uniform mass flux $j$ by:
\begin{equation}
\delta^2 = 1-\frac{v_\infty^2}{c_\infty^2}=1 - \left [ \frac{j}{\pi \hbar \,n_\infty^2}\right ]^2 \, ,
\label{delta}
\end{equation}

where $v_\infty$ and $c_\infty$ are the fluid and sound velocity at $|x|\to\infty$ respectively. The solution is then parametrized by the asymptotic density $n_\infty$ and the mass flux $j$. The asymptotic density $n_\infty$ defines the length units for both the density profile $n(x)$ and the coordinate $x$, showing that, at fixed $n_\infty$, Eqs. (\ref{solidens},\ref{soliphase}) provide a family of solutions which only depends on the dimensionless parameter $0\le \delta \le 1$ (\ref{delta}) or, equivalently, on the density drop 

\begin{equation}
\bar n_0=\frac{n(0)}{n_\infty} = \sqrt{4-3\delta^2}-1 \, .
\label{depletion}
\end{equation}

With the adopted choice of signs the fluid is left moving (i.e., $j < 0$). The chemical potential is given by $\mu=\frac{\hbar^2\pi^2(2-\delta^2)n_\infty^2}{2m}$, which equals the chemical potential of the uniform solution at the same asymptotic density $n_\infty$ and conserved current $j$. Therefore, the soliton represents an excited state with finite excitation energy, as in the weakly interacting case \cite{pita}. 

Similarly, in the co-moving frame, a many-body soliton should be an exact, stationary solution of the many-particle Schr\"odinger equation, breaking the translational symmetry of the Hamiltonian. In the case of a TG gas, it corresponds to a symmetry breaking eigenstate of the free Fermi gas. A hint towards the identification of such a many-body state comes from the observation that a \textit{half soliton} \cite{pavloff} of the (cubic or quintic) GPE is precisely a steady state solution of a Bose gas in an external step potential of height $V_0$ \cite{universe}. The particular \textit{half soliton} wavefunction has the unique feature of displaying a flat density profile beyond the step and requires a well-defined relation between the step height $V_0$ and the asymptotic density $n_\infty$:
\begin{equation}
\left (\frac{2\pi\,n_\infty}{Q} \right )^2= \frac{18\,(1-\delta^2)}
{(4-3\,\delta^2)^2 - \sqrt{4-3\,\delta^2}\,(8-9\,\delta^2)}
\label{nq}
\end{equation}
with $\hbar Q=\sqrt{2mV_0}$. Analogously, the full soliton can be obtained by considering, instead of the step potential, a square well of depth $-V_0$ and width $a$ and choosing the stationary solution with a flat density profile inside the well. This condition requires the same relation (\ref{nq}). Then, by letting the well width $a\to 0$ at fixed $V_0$, the free solitonic solution (\ref{psi}) is recovered. 

\begin{figure} [ht]
\includegraphics[width=8cm]{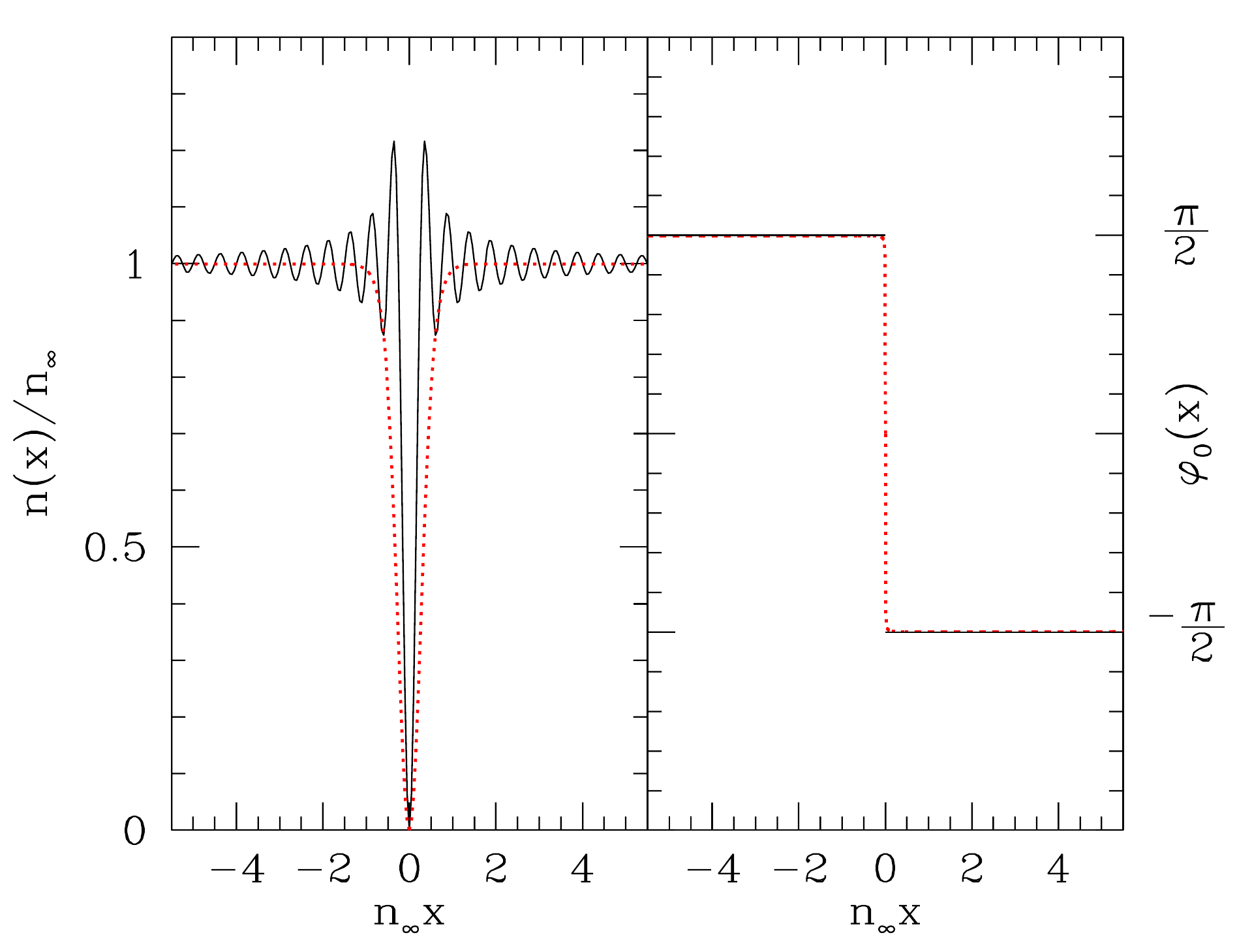}
\caption{Dark soliton solution for a strongly interacting Bose gas in the limit of vanishing current: $j\to 0$. The red dotted lines represent the quintic GPE solution (\ref{psi}) while the black solid ones the exact TG stationary eigenstate (\ref{manybody}). Left panel: density profile; right panel: phase shift in the laboratory frame.}
\label{compara1}
\end{figure}

\begin{figure} [ht]
\includegraphics[width=8cm]{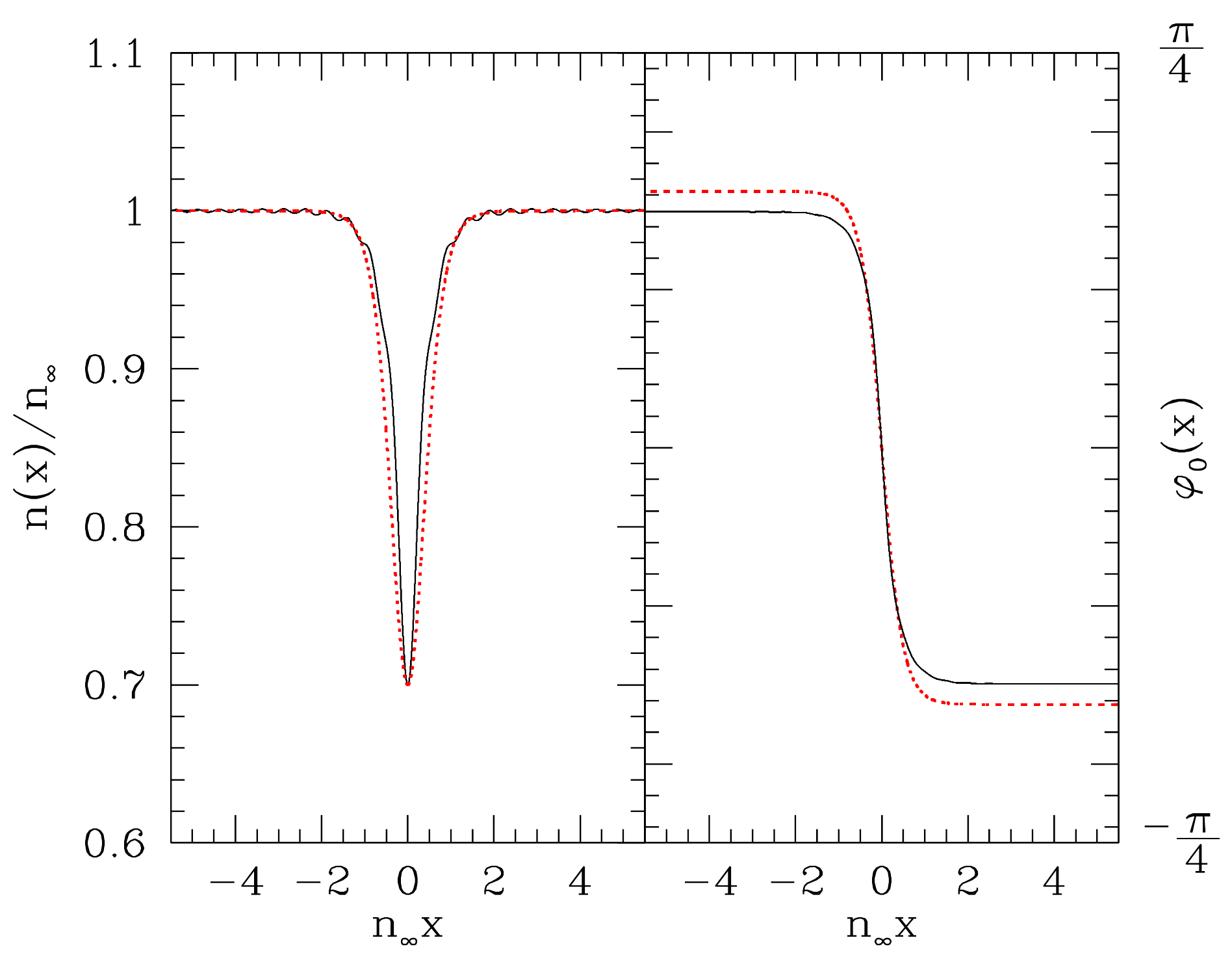}
\caption{Grey soliton solution for a strongly interacting Bose gas for $\bar n_0=0.7$. The red dotted lines represent the quintic GPE solution (\ref{psi}) while the black solid ones the exact TG stationary eigenstate (\ref{manybody}). Left panel: density profile; right panel: phase shift in the laboratory frame. 
}
\label{compara2}
\end{figure}

This procedure can be easily generalized to the free Fermi gas. Remarkably, it is possible to find a set of exact single-particle eigenstates of the square well potential characterized by a flat density profile inside the well, in close analogy
to the relevant solution of the GPE equation previously identified. These states are suitable linear combinations of the scattering eigenfunctions of positive and negative momentum and, in the $a\to 0$ limit, they are simply given by: 
\begin{equation}
\psi_k(x) = \frac{1}{\sqrt{2\pi\, \left [ 1+R_k^2\right ]}}\,\left [ e^{ikx}+R_k\,e^{-ikx}\right ] 
\label{singlepsi}
\end{equation}
where $k \le 0$, $p=\sqrt{k^2+Q^2}$ and $R_k=\frac{k+p}{k-p}$. 
These combinations have a well-defined single-particle energy $\epsilon_k = \frac{\hbar^2 k^2}{2m}$ and thus represent stationary states. In the co-moving frame the many-body exact eigenstate can be written as a Slater determinant 
of these single particle orbitals for momenta in the interval $-2k_F \le k \le 0$: 
\begin{equation}
\Psi_B(x_1,...,x_N)=\left[ \prod_{j>i} \text{sgn}(x_j-x_i) \right] \text{det} \, \psi_k(x_i) \, ,
\label{manybody}
\end{equation}
with $\psi_k(x_i)$ given by Eq. (\ref{singlepsi}). In this case, the asymptotic density is just $n_\infty= \frac{k_F}{\pi}$ while the mass flux is
\begin{equation}
j = - \frac{\hbar Q^2}{8\pi} \,\left [ \chi - \arctan \chi\right ] \, ,
\end{equation}
with $\chi =2\,\frac{2k_F}{Q}\sqrt{\left (\frac{2k_F}{Q}\right)^2 +1} $. 
The full density profile of this many-body eigenstate is not homogeneous and is analytically given by
\begin{equation}
n(x) = \int_{-2k_F}^0 \frac{dk}{2\pi} \,\left [ 1 +\frac{2\,R_k}{1+R^2_k}\,\cos(2kx) \right ] \, ,
\label{prof}
\end{equation}
while the density drop is 
\begin{equation}
\bar n_0=\frac{n(0)}{n_\infty} = 1 - \frac{Q}{2\sqrt{2}k_F} \,\arctan \frac{2\sqrt{2}k_F}{Q} \, .
\label{deplex}
\end{equation}
In summary: starting from a strongly interacting, flowing Bose gas in presence of an external potential well of given parameters $(Q,a)$ we can find the stationary states with a constant density inside the well either by solving the quintic GPE or by use of the exact Bose-Fermi mapping. Each class of solutions depends on a single dimensionless quantity, which can be conveniently identified as the density 
at infinity $\frac{n_\infty}{Q}=\frac{k_F}{\pi\,Q}$. 
The mass current, the density drop and the full density profile are expressed in terms of such a parameter. 
Then, by taking the $a\to 0$ limit the square well disappears and the dimensional parameter $Q$ defining the well 
depth loses its physical meaning. Nevertheless, the density profile remains non uniform in both cases and coincides with that of the the GPE soliton (\ref{solidens}) and of the TG stationary state (\ref{prof}), respectively. Both states depend only on one dimensional and one dimensionless parameter: the asymptotic density $n_\infty$ and the density drop $\bar n_0$ Eqs. (\ref{depletion},\ref{deplex}).  A comparison between the two profiles is shown in Figs. \ref{compara1} and \ref{compara2}, where the (normalized) density and the phase of the solitonic solution is plotted for two representative choices of the density drop $\bar n_0$ (\ref{depletion},\ref{deplex}). The phase $\varphi(x)$ is defined as the integral of the local velocity: $\varphi(x) = \int_0^x dx^\prime \, \frac{j}{\hbar\,n(x^\prime)}$ and to better appreciate the phase shift induced by the soliton, the figures display the phase of the wavefunction in the laboratory frame, where the fluid at infinity is at rest: $\varphi_0(x) = \varphi(x) - \frac{j}{\hbar\,n_\infty}\,x$. 
The agreement between the mean-field and the exact many-body solution is remarkable for both the density profile and the phase shift. Furthermore, the solutions display all the dark soliton properties: indeed, even in the absence of any external potential, there is a notch in the density profile and a precise phase jump which do not change in time. Moreover, the exact stationary eigenstate and the GPE soliton are stationary solutions characterized by a well-defined constant velocity $v_{sol} = -\frac{j}{n_\infty}$, thus representing constantly moving rarefactions in the laboratory frame. The qualitative difference between the shape of the GPE soliton and the exact result is the presence of an oscillating power-law tail slowly approaching the asymptotic density in the TG solution, while the GPE soliton is exponentially localized \cite{note}.
Remarkably, in the limit $\frac{k_F}{Q}\to 0$ (Fig. \ref{compara1}) , our expression reduces to 
the proposal put forward in Ref \cite{tgsol1}.

\section{Experimental realization} The next question we want to address is whether the previously identified solitonic solutions can be generated by means of some external perturbation applied on a homogeneous, flowing BEC, in order to set the system out of equilibrium, consequentially triggering the formation of stable structures. In the laboratory, two alternative methods are currently adopted to create solitonic structures: \textit{phase imprinting} and \textit{density engineering} \cite{review}. In both cases, the starting configuration is a uniform (possibly flowing) BEC described by the trivial solution of the GPE equation $\Phi(x) = \sqrt{n_\infty}\,e^{-ik_0x}$ or, in the Fermi representation, by the Slater determinant of plane waves $\psi_k(x)= \frac{e^{ikx}}{\sqrt{2\pi}}$ with $-k_F-k_0 \le k \le k_F-k_0$, where, in both cases, $k_0$ is related to the fluid velocity by $v=\frac{\hbar k_0}{m}$. 

\subsection{Phase imprinting}

\begin{figure} [ht]
\includegraphics[width=8cm]{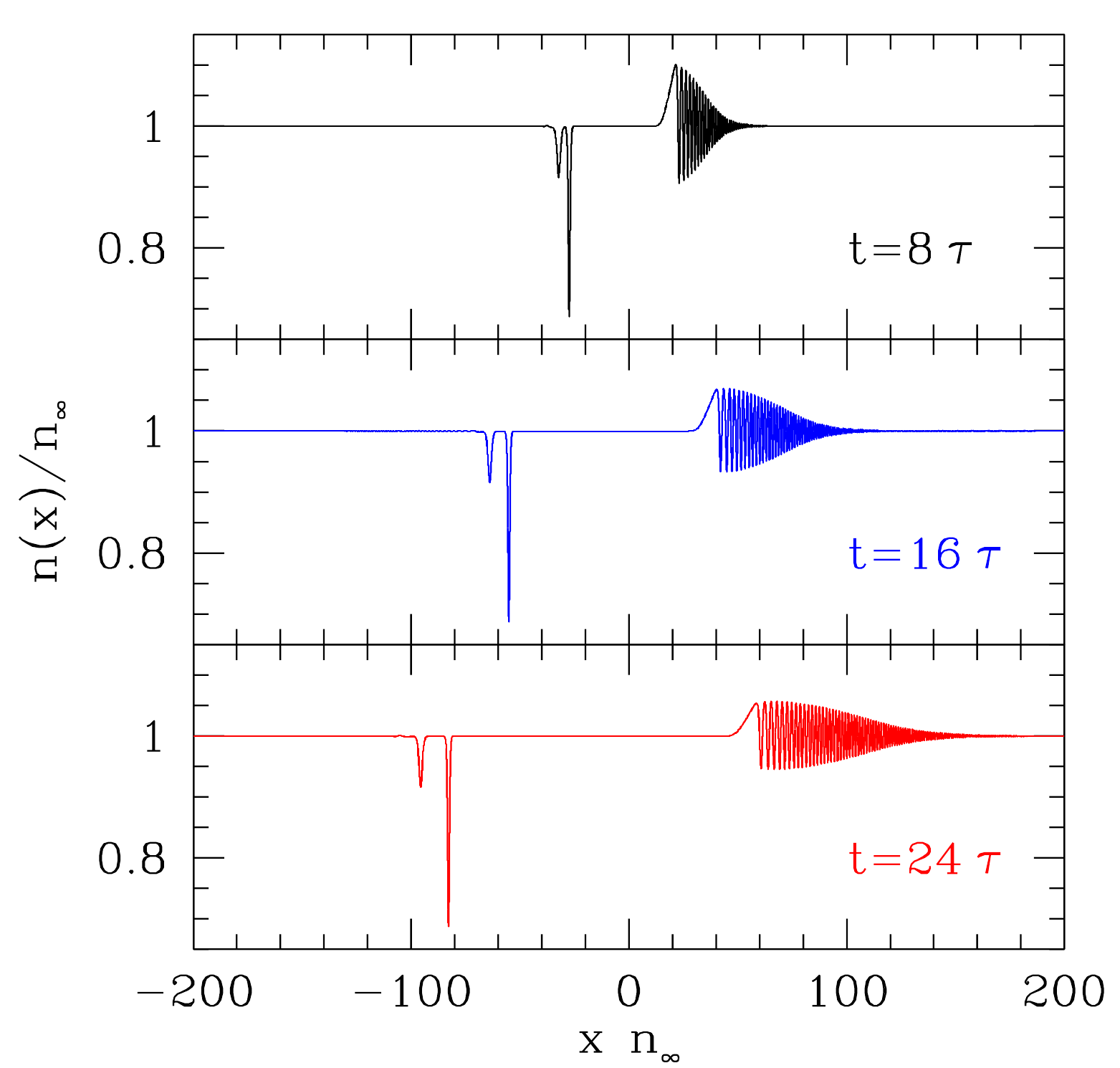}
\caption{Time evolution of the condensate density after phase imprinting. The three curves correspond to different dimensionless times with $\tau=\frac{m}{\hbar \,n_\infty^2}$.}
\label{phase}
\end{figure}

For the phase imprinting procedure, the condensate wavefunction is initially perturbed by a non-uniform phase change of the form: 
\begin{equation}
\Phi(x) = \sqrt{n_\infty} \,\,e^{-i(k_0x-\frac{\pi}{2} \tanh \alpha x)}
\end{equation}
(where $\alpha$ is a tuning parameter) and then the cloud is let evolve freely. The time evolution according to the quintic GPE 
is shown in Fig. \ref{phase} for $k_0=n_\infty$ and $\alpha=n_\infty$. 

\begin{figure} [ht]
\includegraphics[width=8cm]{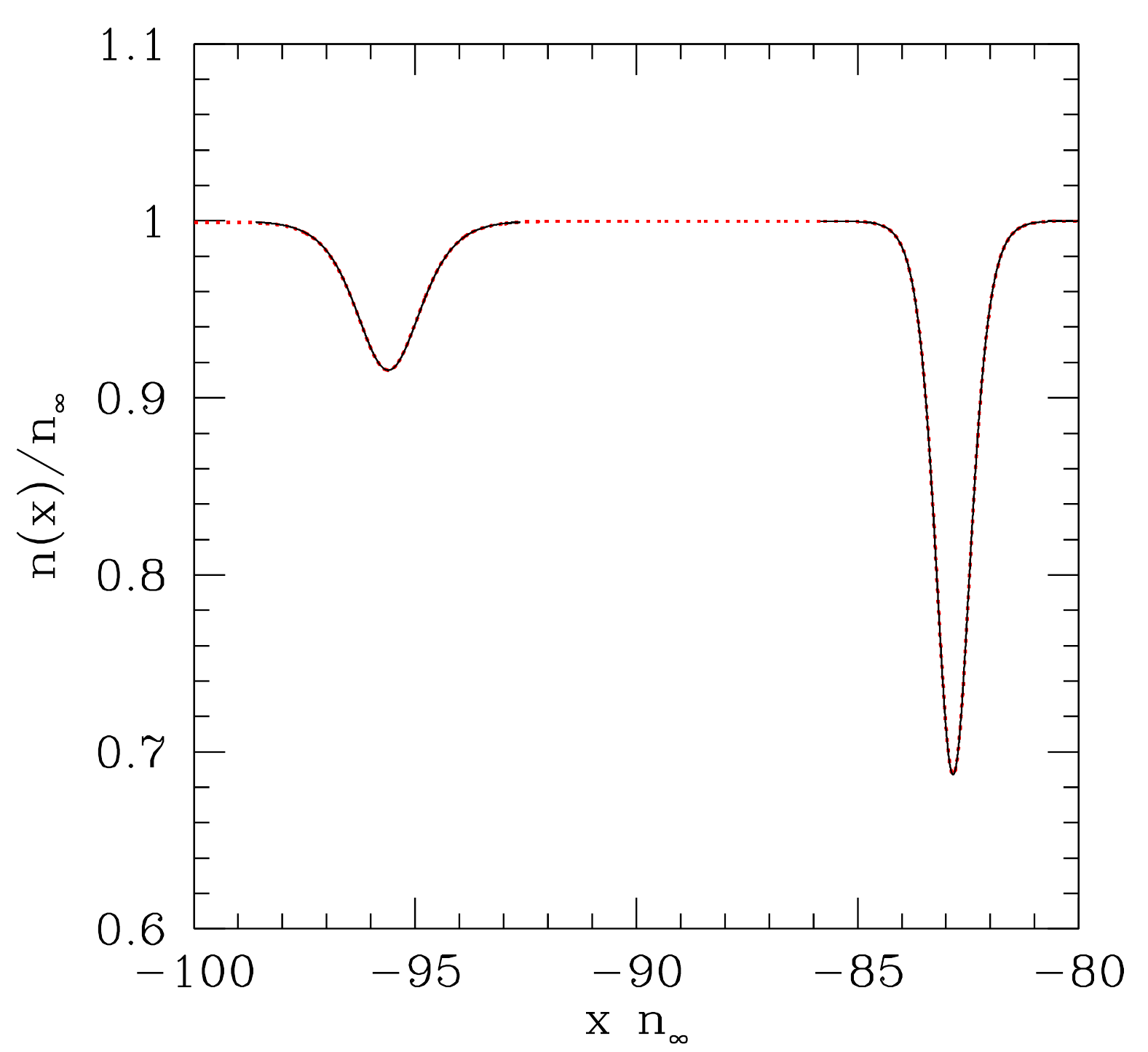}
\caption{Comparison between the shape of the two soliton-like structures at time $t=24\,\tau$ (red dotted lines) 
and the analytical solitonic solutions Eq. (\ref{solidens}) (black solid lines) for $\bar n_0=0.9156$ (left curve) and 
$\bar n_0= 0.6872$ (right curve). The analytical solitons have been shifted to match the position of the minima of the two numerical dips. The numerical and the analytical form are perfectly superimposed and the agreement is remarkable.}
\label{zoom}
\end{figure}

The presence of a propagating dispersive shock wave (DSW) on the right is evident, while on the left two dark soliton-like structures appear, moving at different velocities. In Fig. \ref{zoom} we enlarge the soliton region and we superimpose the analytical form previously obtained, for two different values of the density drop. The good agreement confirms the solitonic nature of the sharp features shown in Fig. \ref{phase}.

\begin{figure} [ht]
\includegraphics[width=8cm]{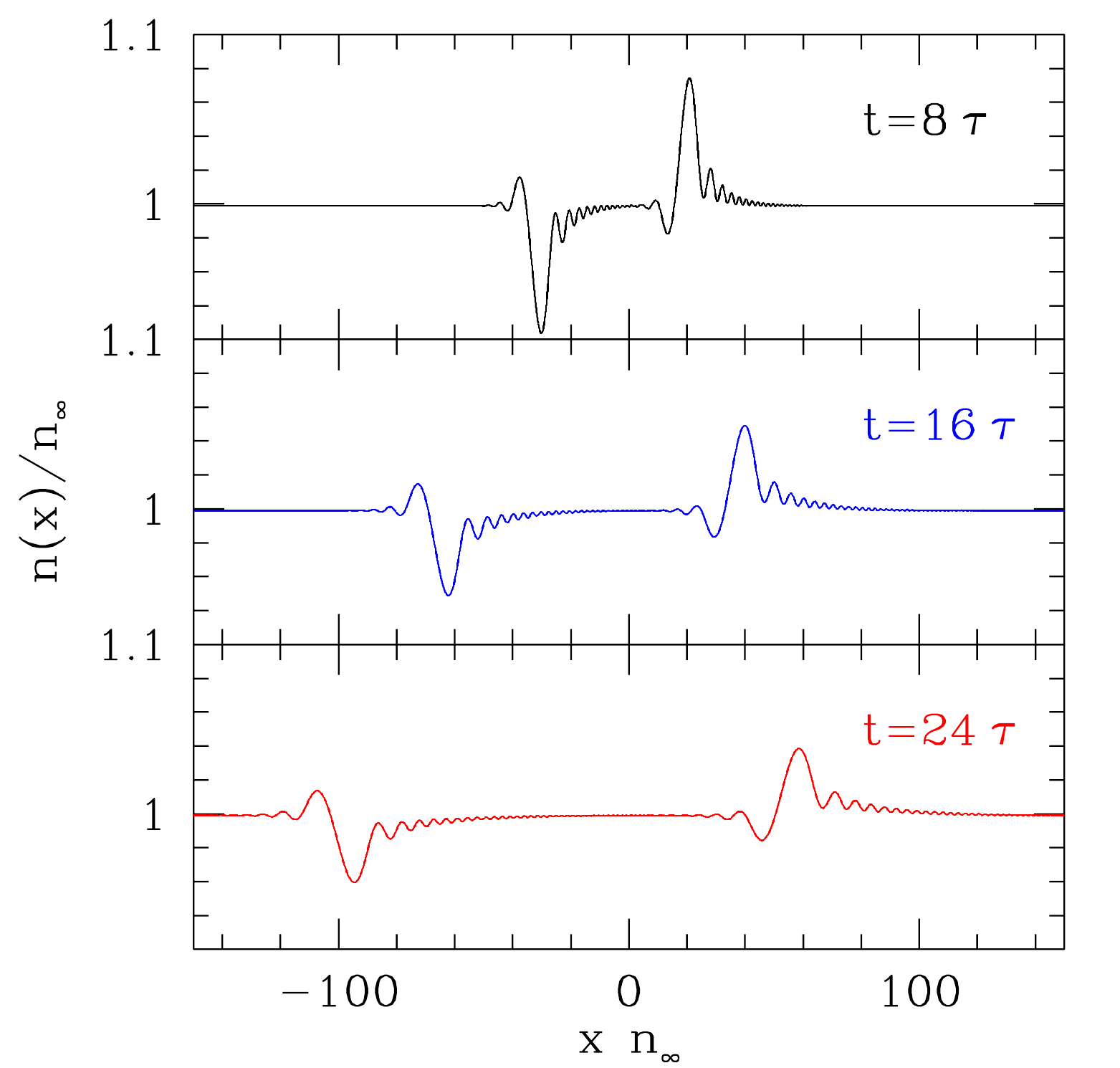}
\caption{Time evolution of the density of a TG gas after phase imprinting. The curves correspond to three different dimensionless times with $\tau=\frac{m}{\hbar \,n_\infty^2}$.}
\label{tonksphase}
\end{figure}

The exact dynamics of a Tonks gas after phase imprinting can be evaluated by integrating the Schr\"odinger equation for
a free Fermi gas. The initial condition at $t=0$ is given by a Slater determinant of ``phase imprinted" plane waves with $-k_F-k_0 \le k \le k_F-k_0$:
\begin{equation}
\psi_k(x) = \frac{1}{\sqrt{2\pi}}\,e^{i(kx+\frac{\pi}{2} \tanh \alpha x)} \, .
\end{equation}
A few snapshots of the TG dynamics are shown in Fig. \ref{tonksphase} using the same set of 
parameters $(k_0,\alpha)$ of the GPE case. While similar structures are formed in the first stages of the evolution, 
the dark-soliton-like dip on the left is much smaller and further weakens in time. 
The exact dynamics does not appear to lead to the formation of stable, non-homogeneous structures. 
Moreover, the dispersive shock waves clearly visible in the GPE dynamics - Fig. \ref{phase} - are lacking in the exact TG evolution, being replaced by smooth density peaks which propagate at the same velocity of the DSW.
Different sets of parameters (including smaller values of $\alpha$) have been checked following a previous study \cite{polacchi} suggesting that the dynamics of a TG gas after phase imprinting strongly depends on the length scale $\alpha^{-1}$ governing the initial phase change. In these cases, although the ensuing GPE time evolution is smoother, the results are qualitatively similar to the cases previously illustrated: no solitonic structure forms according to the exact dynamics, while it is always present in mean-field approximation.

\subsection{Density engineering}

The density engineering method is another scheme commonly used in the laboratory to generate solitons. In this case a quench in the density (rather than in the phase) is applied to a uniform condensate by means of a suitable external perturbation. The potential applied in order to obtain a solitary emission can be of different forms \cite{review} and here we choose a square well of width $a$ and depth $V_0=\frac{\hbar^2 Q^2}{2m}$, which is suddenly generated at $t=0$. 
In the following, we will take the atoms to be at rest before the quench (but the same technique can be applied to flowing gases), the initial density is set as $n=\frac{Q}{\pi}$ and the well width given by $Qa=2$. 

\begin{figure} [ht]
\includegraphics[width=8cm]{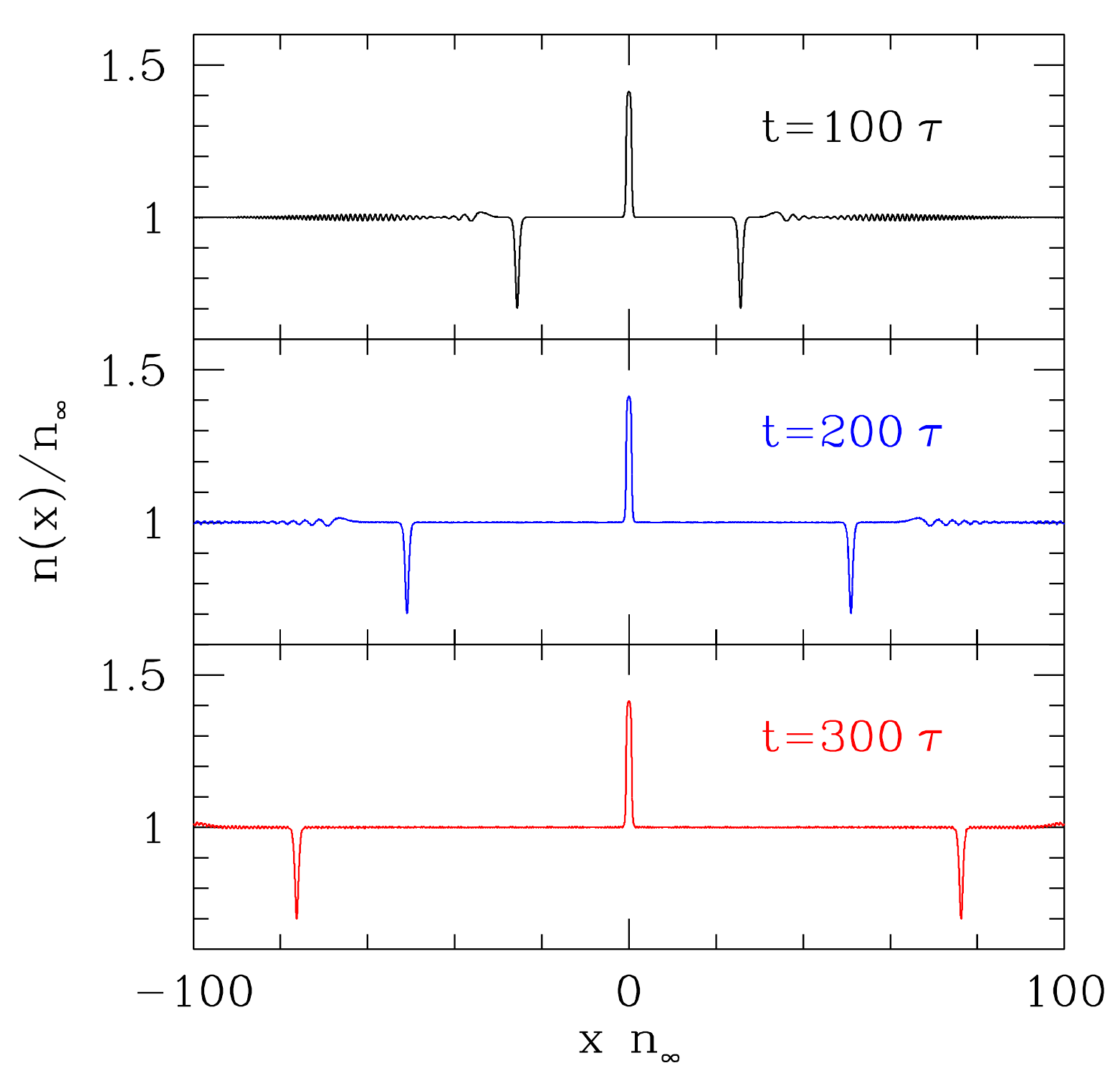}
\caption{Time evolution of the condensate density after the quench for the GPE case. The three curves correspond to different dimensionless times: with $\tau=\frac{m}{\hbar \,Q^2}$.}
\label{eng}
\end{figure}

Fig. \ref{eng} displays a few snapshots of the (quintic) Gross-Pitaevskii equation dynamics at three different times with the aforementioned initial conditions. As the curves show, after the quench the gas is perturbed in the region of the potential 
well and solitons are emitted from both sides, propagating away from the potential. Dispersive shock waves are also evident near the soliton structures.

\begin{figure} [ht]
\includegraphics[width=8cm]{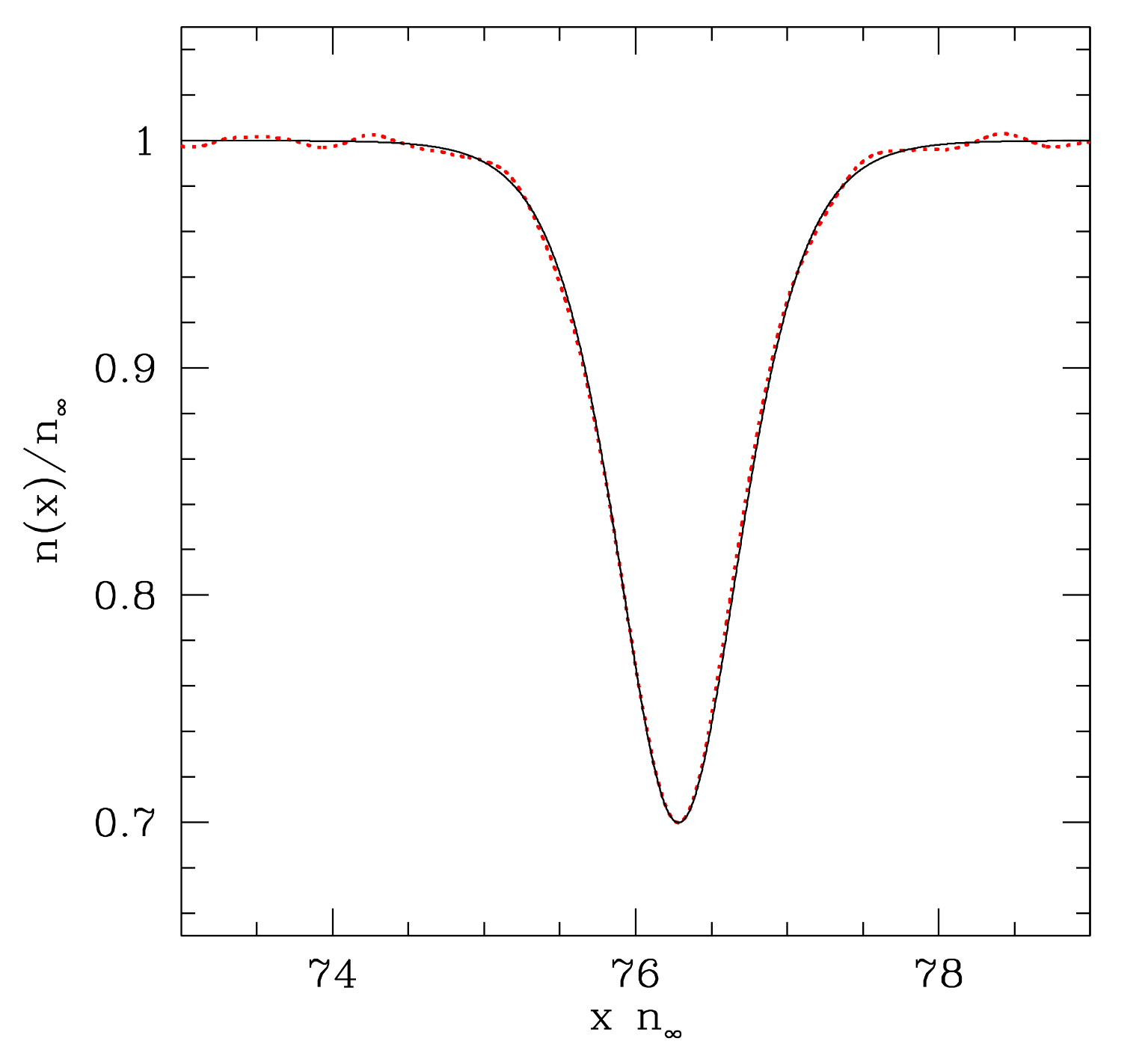}
\caption{Soliton structure of Fig. \ref{eng} at $t=300\,\tau$ (red dotted curve) superimposed with the analytic solution given in the article (black solid curve). The two curves are almost identical and thus difficult to distinguish on the canvas.}
\label{solieng}
\end{figure}

Fig. \ref{solieng} shows an enlargement of the soliton structure present in Fig. \ref{eng} at $t=300\,\tau$, superimposed with the analytic form previously detailed evaluated for the same density drop. The matching between the two curves is remarkable, demonstrating that the dip propagating away from the potential is indeed an actual soliton. 

\begin{figure} [ht]
\includegraphics[width=8cm]{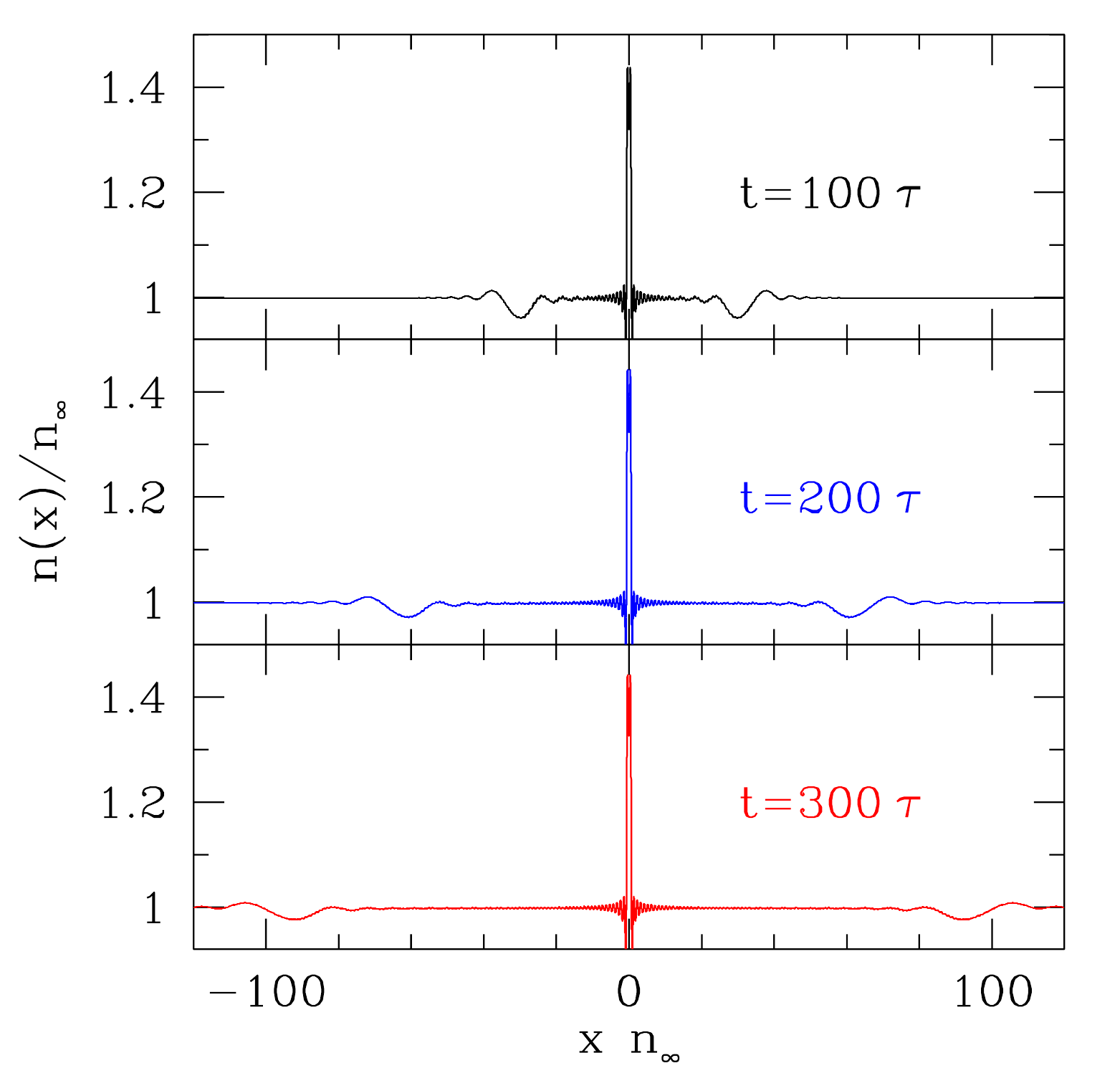}
\caption{Time evolution of the density of a Tonks-Girardeau gas after a quench. The three curves correspond to different dimensionless times with $\tau=\frac{m}{\hbar \,Q^2}$.}
\label{tonkswell}
\end{figure}

\begin{figure} [ht]
\includegraphics[width=8cm]{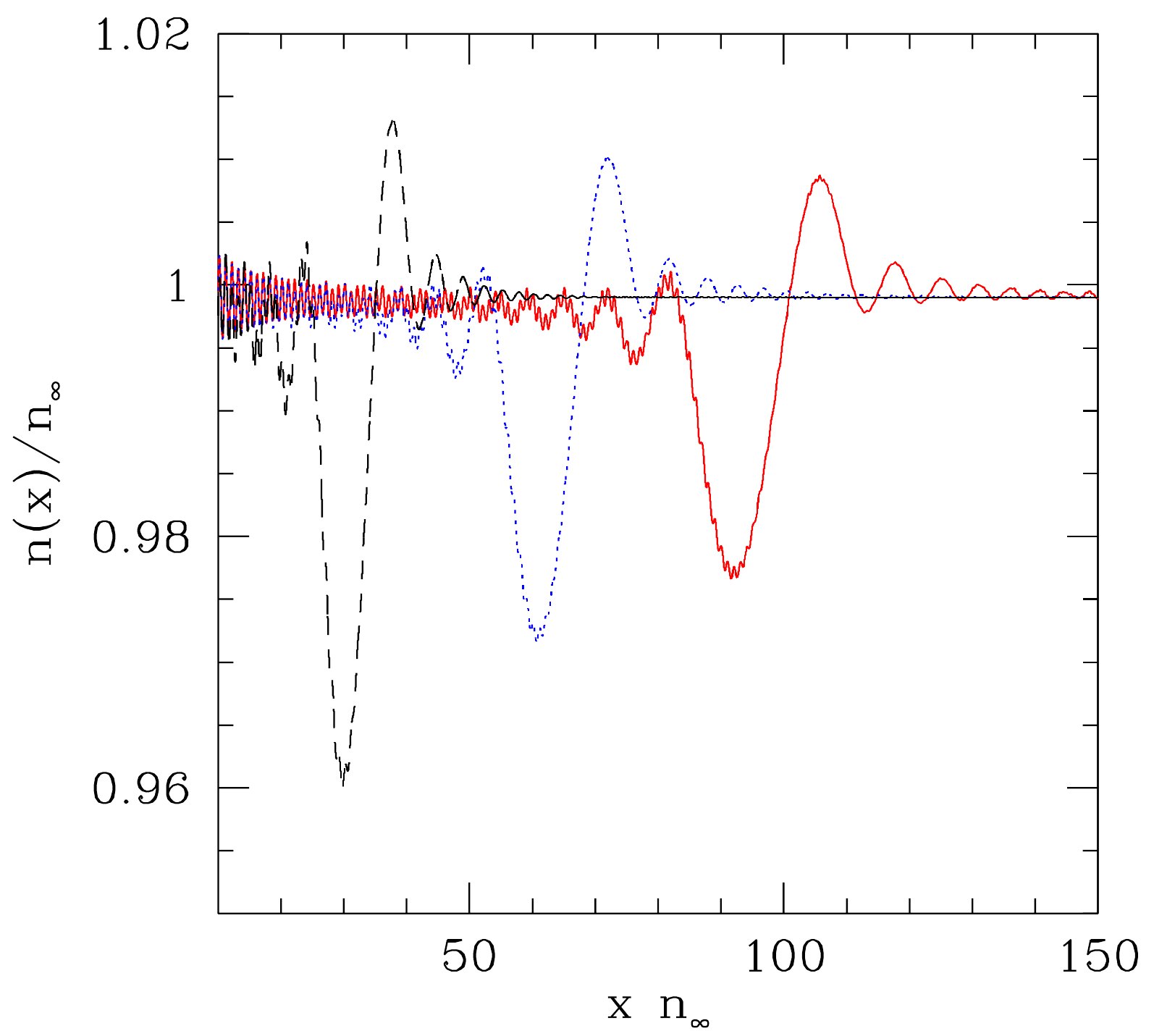}
\caption{An enlargement of Fig. \ref{tonkswell} which shows the rarefaction created in the quench and its evolution in time. 
The three curves correspond to the three times plotted in Fig. \ref{tonkswell} : $t=100\,\tau$ (dashed black); $t=200\,\tau$ (dotted blue); $t=300\,\tau$ (solid red) with $\tau=\frac{m}{\hbar \,Q^2}$.}
\label{tonkswellzoomed}
\end{figure}

We now compare the results to the exact Tonks-Girardeau (TG) dynamics under the same conditions. Now the initial state at $t=0$ is a Slater determinant of plane waves and the time evolution is governed by the Schr\"odinger equation with the same potential used in the GPE dynamics. A few snapshots of the TG exact evolution are given in Figs. \ref{tonkswell} and \ref{tonkswellzoomed} and we can see that the dynamics is qualitatively the same: a perturbation appears in the region of the potential well while density dips propagate in both directions, analogously to the GPE case. Differently from the mean-field case, though, these structures do not behave as solitons as they suffer considerable dispersion upon evolution. Furthermore, DSW are absent. 
Thus we can conclude that, as for the phase imprinting case, density engineering is a valid technique to generate 
solitons at mean-field level but the same cannot be said for the many-body solution of the TG gas. 

Finally, the proposal put forward in \cite{deltaosc} has been studied and a time-dependent potential has been 
applied to the gas in order to excite the correct eigenstates; numerical simulations, however, show that not even 
in this case a stable soliton solution for the TG gas is reached.

\section{Conclusions}

In summary, taking advantage of the known Bose-Fermi mapping in one dimension, an exact, stationary solution of the many-body problem for a Tonks-Girardeau gas in one dimension showing the same features attributed to a dark soliton has been found. This solution has a power-law oscillating decay and describes a density rarefaction which moves with constant speed (lower than the sound speed) in a uniform TG gas. Both the density and the phase profile resemble the analytical solitonic solution of the (quintic) GPE. 
The GPE evolution after both phase imprinting and quench gives rise to soliton trains extremely well represented by 
the analytical solutions we found, at variance with the exact dynamics of a TG gas under the same initial conditions, 
which shows no evidence of solitonic structures. The absence of a dynamical mechanism to spontaneously generate 
solitons in a TG gas could be explained by recalling the analytical results of Ref. \cite{prd}, where it was shown that 
the long time dynamics of a Tonks-Girardeau gas always leads to the stationary state built with the exact single-particle scattering states of the external potential. Although the soliton-like wavefunction we have introduced in Eq. (\ref{manybody}) is indeed a Slater determinant of suitable single particle orbitals (\ref{singlepsi}), they cannot be identified as pure scattering states. 

This investigation has been specifically performed for the TG gas, whose properties are known to represent 
the generic behavior of one-dimensional, interacting Bose gas. However, the adopted procedure can be 
straightforwardly generalized to any one-dimensional fluid: The solitonic wavefunction is obtained starting 
from the exact symmetrical eigenstates of the interacting Bose gas in an external potential 
well and then letting the well width to zero.


\begin{thebibliography}{100}
\bibitem{note} These types of oscillations are expected in the case of fermions \cite{antezza}.
\end{thebibliography}
\end{document}